\documentclass[sigconf]{acmart}
\AtBeginDocument{%
  }

\setcopyright{acmlicensed}

\copyrightyear{2025}
\acmYear{2025}
\setcopyright{acmlicensed}\acmConference[MICRO '25]{58th IEEE/ACM
International Symposium on Microarchitecture}{October 18--22, 2025}{Seoul,
Republic of Korea}
\acmBooktitle{58th IEEE/ACM International Symposium on Microarchitecture
(MICRO '25), October 18--22, 2025, Seoul, Republic of Korea}
\acmDOI{10.1145/3725843.3756071}
\acmISBN{979-8-4007-1573-0/2025/10}

\usepackage{multirow}

\usepackage{xcolor}   
\usepackage{framed}   

\colorlet{rqbg}{gray!20} 
\newenvironment{researchquestionbox}{%
  \MakeFramed{\advance\hsize-\width \FrameRestore}%
  \noindent
}{%
  \endMakeFramed
}

\newcommand{\researchquestion}[1]{%
  \begin{researchquestionbox}#1\end{researchquestionbox}%
}

\begin{document}

\title{Kelle: Co-design KV Caching and eDRAM for Efficient LLM Serving in Edge Computing}
\author{Tianhua Xia}
\affiliation{%
  \department{Tandon School of Engineering}
  \institution{New York University}
  \city{New York}
  \state{NY}
  \country{USA}
}
\email{tx856@nyu.edu}

\author{Sai Qian Zhang}
\affiliation{%
  \department{Tandon School of Engineering}
  \institution{New York University}
  \city{New York}
  \state{NY}
  \country{USA}
}
\email{sai.zhang@nyu.edu}




\begin{abstract}
  Running Large Language Models (LLMs) on edge devices is crucial for reducing latency, improving real-time processing, and enhancing privacy. By performing inference directly on the device, data does not need to be sent to the cloud, ensuring faster responses and reducing reliance on network connectivity. However, implementing LLMs on edge devices presents challenges, particularly with managing key-value (KV) caches, which plays a pivotal role in LLM serving. As the input text lengthens, the size of the KV cache increases linearly with the sequence length, leading to a significant memory footprint and data access costs. On the other hand, edge devices have limited memory and computational power, making it hard to store and efficiently access the large caches needed for LLM inference.

  To mitigate the substantial overhead caused by KV cache, we propose using embedded DRAM (eDRAM) as the primary storage for LLM serving in edge device, which offers higher storage density compared to SRAM. However, to ensure data integrity, eDRAM needs periodic refresh operations, which are power-intensive. To reduce eDRAM costs and improve overall system performance, we propose~\textit{Kelle}, a software-hardware co-design solution optimized for deploying LLMs on eDRAM-based edge systems. Combined with our fine-grained memory eviction, recomputation, and refresh control algorithms, the \textit{Kelle} accelerator delivers a $3.9\times$ speedup and $4.5\times$ energy savings compared to existing baseline solutions.

\end{abstract}


\maketitle

\section{Introduction}
\label{sec:intro}
Large Language Models (LLMs) have demonstrated remarkable capabilities across a wide range of domains. While cloud-based deployments offer advantages like increased processing power, they also come with limitations, including high communication latency and security risks. As LLMs continue to evolve, it is increasingly important to bring their capabilities directly to edge devices~\cite{shi2016edge}. The integration of LLMs into edge devices not only broadens their accessibility but also ensures robust, customized experiences tailored to individual and industrial needs. This trend is gaining traction not only in academia~\cite{frantar2022gptq, 10818760, cai2024edge, yu2024edge, yu2024cambricon}, but also in the industry, where leading companies like Intel~\cite{shen2023efficientllminferencecpus}, NVIDIA~\cite{nvidia-gpu}, Microsoft~\cite{wang2023bitnetscaling1bittransformers}, and Qualcomm~\cite{soriaga2023accelerating} are actively exploring similar solutions.

However, implementing LLMs on edge devices presents challenges, particularly with managing key-value (KV) caches~\cite{radford2019language}, which play a critical role to enhance LLM token generation speed. This mechanism involves storing previously computed~\textit{Key and Value} vectors (KV vectors) during attention calculations and reusing them for generating subsequent tokens. By doing so, it avoids recalculating vectors for earlier tokens with each new token generation. On the other hand, the KV caching incurs a significant memory footprint that grows rapidly as both the model size and the length of generated text increase~\cite{hooper2024kvquant10millioncontext, zhao2024alisa}. For example, when LLaMA 2-7B processes a sequence with the length of 8192 in FP16, the KV cache consumes 4GB of memory, causing the total execution latency to be primarily limited by frequent memory access between on-chip SRAM and off-chip DRAM~\cite{pope2023efficiently, yu2024cambricon}. This becomes particularly problematic in resource-constrained systems with limited on-chip SRAM capacity, such as edge devices~\cite{zhao2024alisa}. For example, the Jetson Orin NX edge GPU has only 4MB L3 cache~\cite{nvidia-gpu}.

A simple way to address this issue is by increasing the on-chip SRAM size, which effectively reduces costly off-chip memory access and enhances overall system performance~\cite{tu2018rana, chen2014dadiannao}. However, edge devices have limited area and power budgets, and expanding SRAM reduces the resources available for other critical components, such as computational cores~\cite{kung1986memory, wang2022exploration, chen2019deep}. 
Alternatively, this study explores the use of embedded DRAM (eDRAM) as the primary on-chip storage medium for KV vectors during LLM execution. With fewer transistors per memory cell, such as 3T for eDRAM cells compared to 6T for SRAM cells, eDRAM provides a higher data storage density, resulting in more than twice the capacity~\cite{giterman20201, chen2014dadiannao}. This increased storage density enables a greater on-chip storage capacity within the same chip area. Moreover, eDRAM also offers much lower leakage power than SRAM (around $3.5\times$ according to prior work~\cite{chun2011edram}). These benefits make eDRAM an attractive option for storing KV vectors in edge devices. 

\begin{figure}
    \centering
    \includegraphics[width=0.95\columnwidth]{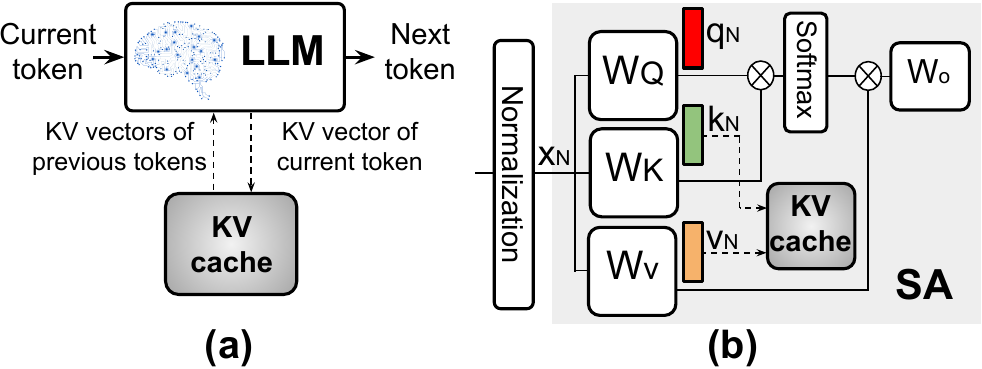}
    \caption{(a) LLM token generation. (b) KV cache for intermediate data storage, where N denotes the token index.}
    \label{fig:llm-intro2}
\end{figure}

However, a key drawback of eDRAM is the need for periodic refreshes to prevent data loss from leakage. Specifically, refreshing eDRAM cells requires a read-write operation, increasing latency and power consumption, which can significantly impact the efficient deployment of LLMs. To address the challenges of integrating eDRAM, we co-design the KV caching algorithm with the eDRAM-based hardware system, enabling a highly efficient KV cache implementation without compromising accuracy. Our contributions are summarized:

\begin{itemize}
\item We propose~\textit{Kelle}, an algorithm-system co-design solution for in-device LLM serving on eDRAM-based edge systems. 
To optimize eDRAM integration cost and improve LLM execution efficiency, we introduce~\textit{attention-based eviction and recomputation policy} (AERP) and~\textit{two-dimensional adaptive refresh policy} (2DRP) for efficient KV cache implementation (Section~\ref{sec:aep} and~\ref{sec:2drp}).
\item We design a~\textit{Kelle accelerator} that utilizes eDRAM as the primary on-chip storage, featuring a customized memory layout. To maximize efficiency, the accelerator integrates a dedicated eDRAM controller and a~\textit{systolic evictor} for efficient AERP and 2DRP implementation (Section~\ref{sec:Kelle-hardware}).
\item We also introduce the Kelle scheduler (Section~\ref{sec:Kelle-compiler}), which adopts an efficient computation pattern to optimize eDRAM data lifetime and LLM serving latency, significantly reducing both eDRAM refresh energy and memory traffic.
\item The evaluation results show that Kelle achieves a $3.9\times$ speedup and $4.5\times$ energy savings compared to other baseline hardware platforms, while maintaining a negligible impact on LLM accuracy (Section~\ref{sec:accuracy-eval},~\ref{sec:hardware-eval}).
\end{itemize}

\section {Background and Related Work}
\label{sec:background}
\subsection{LLM Workflow}
\label{sec:bg:workflow}
Modern LLMs (e.g., Llama series~\cite{touvron2022llama,touvron2023llama}, GPT series~\cite{radford2019language,brown2020language}) are constructed as a stack of transformer decoders, with each decoder comprising two fundamental components: a Self-Attention (SA) block and a feed forward network (FFN). During the LLM serving process, the input of the SA block is first multiplied with three weight matrices $W_{Q}$, $W_{K}$, and $W_{V}$, yielding the outputs termed query ($q$), key ($k$), and value ($v$), respectively. 
The resulting $q$ and $k$, in combination with $v$, will then undergo multiplication, softmax, and residual addition to generate the SA output.
The output from the SA will then be passed to the FFN for further processing, which typically involves a standard MLP~\cite{radford2018improving, radford2019language} or gated MLP~\cite{liu2021pay, touvron2022llama, touvron2023llama}. The FFN consists of multiple fully connected (FC) layers along with an intermediate activation function, such as GeLU~\cite{hendrycks2016gaussian}. 

LLM serving involves two main stages: pre-filling and decoding. In the pre-filling stage, the model processes the context tokens in parallel. During the decoding stage, the model predicts the next token based on the current and previous tokens. This is done by combining the current input with information from previous tokens, expressed in terms of their Key and Value (KV) vectors. 
This process is repeated in an auto-regressive manner (Figure~\ref{fig:llm-intro2} (a)). 

\subsection{KV Caching}
\label{sec:kv-background}
\begin{figure}
    \centering
    \includegraphics[width=0.95\columnwidth]{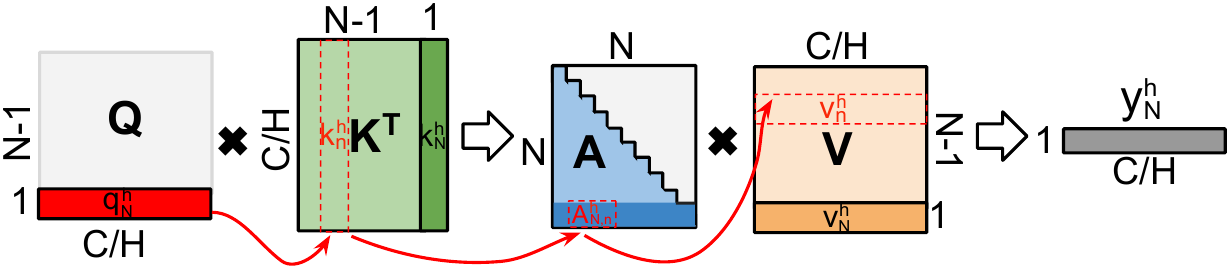}
    \caption{An example on KV vector computation.}
    \label{fig:kv_compute}
\end{figure}

During the decoding stage, the KV vectors of each newly generated token are stored in a KV cache to enhance generation speed, as shown in Figure~\ref{fig:llm-intro2} (b). By doing so, it avoids recalculating vectors for earlier tokens with each new token generation. Specifically, to produce the output of an LLM block, the input vector of Nth token, with a length \( C \), is multiplied by \( W_{Q} \), \( W_{K} \), and \( W_{V} \) to generate three vectors: query \( q_{N} \), key \( k_{N} \), and value \( v_{N} \), each with dimensions \( 1 \times C \), where \( C \) denotes the channel size, as shown in Figure~\ref{fig:llm-intro2} (b). $q_{N}$, together with other KV vectors are then split along the channel dimension into multiple parts, each has a dimension of $1\times \frac{C}{H}$, where $H$ denotes number of heads. The resulting vectors for head h are denoted as $q^{h}_{N}$, $k^{h}_{N}$, $v^{h}_{N}$, respectively. The KV vectors from the previous $N-1$ tokens are then loaded from memory. For each head h, the dot products are then performed between $q^{h}_{N}$ and each of the key vectors $k^{h}_{n}, 1 \leq n \leq N$, and the result is passed through a softmax function, yielding a vector of attention scores, denoted as $A^{h}_{N}$, with dimensions of $1 \times N$. Next, the attention score vector is used to compute a dot product with each of the value vectors $v^{h}_{n}, 1 \leq n \leq N$, producing a result vector $y^{h}_{N}$ of length $\frac{C}{H}$. $y^{h}_{N}$ will then be concatenated across multiple heads, the resulting vector $y_{N}$ which is further multiplied with $W_{O}$. The process is illustrated in Figure~\ref{fig:kv_compute} and can be represented by the following equations:
\begin{equation}
\label{eqn:kv-computation1}
A^{h}_{N} = softmax([q^{h^{\top}}_{N}k^{h}_{1},q^{h^{\top}}_{N}k^{h}_{2},...,q^{h^{\top}}_{N}k^{h}_{N}])
\end{equation}

\begin{equation}
\label{eqn:kv-computation2}
y^{h}_{N} = \sum_{1\leq n\leq N}(A^{h}_{N,n} \cdot v^{h}_{n})
\end{equation}

where $A^{h}_{N,n}$ represents the $n$-th element of $A^{h}_{N}$. As per Equation~\ref{eqn:kv-computation1} and Equation~\ref{eqn:kv-computation2}, we notice that the relative order of KV vector pairs $[k^{h}_{n}, v^{h}_{n}]$ does not affect the decoding computation. In other words, if we swap the values of two pairs of KV vectors (e.g., swap \( [k^{h}_{1}, v^{h}_{1}] \) with \( [k^{h}_{2}, v^{h}_{2}] \)), the result \( y^{h}_{N} \) produced using Equation~\ref{eqn:kv-computation1} and~\ref{eqn:kv-computation2} remains unchanged. 

KV cache compression techniques can be broadly classified into two approaches: token dropping \cite{xiao2024efficientstreaminglanguagemodels, Zhang2023H2OHO, Ge2023ModelTY, Liu2023ScissorhandsET,yang2024pyramidinfer,liu2024minicache} and KV cache quantization \cite{Liu2024KIVIAT, hooper2024kvquant10millioncontext,xiang2024dfrot}. The token dropping strategy identifies and permanently discards unimportant tokens, making them inaccessible thereafter. StreamLLM \cite{xiao2024efficientstreaminglanguagemodels} identifies \textit{sink tokens}, which are tokens at the start of a sequence that are critical for LLM performance, and preserves recent tokens to maintain performance.
H2O \cite{Zhang2023H2OHO} identifies~\textit{heavy hitter tokens} that have high accumulated attention scores. 
KIVI~\cite{KIVI} groups the KV vectors channel-wise to achieve 2-bit asymmetric quantization. QuaRot~\cite{ashkboos2024quarot} utilizes zero-shot Hadamard transformation to reduce the outliers in the model and enables 4-bit quantization. Speculative decoding~\cite{mcdanel2025pipespec,hu2025speculative,hu2025dream,leviathan2023fast} present another inference technique that accelerates LLMs by using a lightweight draft model to propose multiple tokens, which are then selectively verified by the full model. Kelle can be adopted in orthogonal with the speculative decoding techniques.


\subsection{Embedded DRAM}
\label{sec:edram_intro}
\begin{table}
    \caption{Comparison of SRAM and eDRAM.}
    \centering
    \resizebox{0.95\columnwidth}{!}{%
    \begin{tabular}{c c c c c c c}
    \hline 
    65nm    & \multirow{2}{*}{Area}   & Access & Access & Leakage & Refresh & Retention \\
    4MB    &                         & Latency & Energy & Power & Energy & Time        \\ \hline
    SRAM &  7.3 $mm^2$             & 2.6ns & 185.9 pJ/byte & 415$mW$ & NA & NA \\ 
    \hline 
    eDRAM & 3.2 $mm^2$             & 1.9ns & 84.8 pJ/byte & 154$mW$ & 1.14 mJ & 45 $\mu s$ \cite{edram_distribute} \\ 
    \hline 
    \end{tabular}
    }
    \label{tab:sram_edram}
\end{table}

Various eDRAM circuit designs~\cite{giterman20201,yu2020logic} have emerged as alternatives to SRAM, with some requiring only two transistors. Among these, 3T-eDRAM stands out, offering over twice the density and reducing static power dissipation by $3.5\times$~\cite{chun2011edram, llc_edram} compared to SRAM. 
Table~\ref{tab:sram_edram} presents a comparison between 3T-eDRAM and SRAM, with results simulated using Destiny~\cite{destiny} on a 65nm technology node. The advantages of eDRAM, including higher storage density, lower access latency and energy, make it an attractive choice for LLM implementation.
Although eDRAM provides several advantages, a serious drawback is the need for periodic refreshes to prevent data corruption caused by charge leakage. As a result, eDRAM is best suited for storing~\textbf{large amounts of transient data}, where frequent refreshes can be avoided. 

Research has explored using eDRAM in accelerator systems to facilitate CNN computation~\cite{chen2014dadiannao, zhang2024camel, tu2018rana, nguyen2019st}, with proposed methods to mitigate refresh power overhead. DaDianNao~\cite{chen2014dadiannao} partitions eDRAM into banks to mitigate refresh failures but does not address the challenges of refresh energy consumption or data retention. RANA~\cite{tu2018rana} injects the bit retention errors during the training process of CNN to mitigate the accuracy drop caused by low refresh frequency. CAMEL~\cite{zhang2024camel} optimizes CNN model architecture to shorten the data lifetime during training. While prior research has shown the efficiency of eDRAM in Convolutional Neural Network (CNN) inference and training, its potential has yet to be explored for LLMs. In contrast, Kelle focuses on minimizing the off-chip memory access of KV cache in LLMs using eDRAM, an area previously unexplored.

\subsection{Edge LLM Accelerator}
\label{sec:llm-accelerator}
To enable deployment of LLMs on edge devices, several studies have proposed methods to improve the accuracy of quantized transformers~\cite{Guo2023OliVeAL,zadeh2020gobo,Lee2024TenderAL,park2018energy,fang2022efficient,lu2020hardware,zhou2022transpim, liu2024cometparticalw4a4kv4llms,dynax}. 
Tender~\cite{Lee2024TenderAL} suggests a hardware-efficient LLM quantization method by making the scale factor a power of two. COMET~\cite{liu2024cometparticalw4a4kv4llms} designs efficient mixed-precision GPU kernels for 4-bit LLM quantization. 
Other works, such as FlexGen~\cite{sheng2023flexgen}, InfiniGen~\cite{lee2024infinigen}, InstInfer~\cite{pan2024instinfer}, LLM.npu~\cite{xu2024fastondevicellminference}, explore the model offloading strategy between on-chip units and main storage for efficient LLM deployment in resource constraint devices. 
Cambricon-LLM~\cite{yu2024cambricon} proposes a chiplet-based hybrid architecture with NPU and a dedicated NAND flash chip to enable efficient on-device inference. 

\begin{figure}
    \centering
    \hspace*{-2\baselineskip}
    \includegraphics[width=0.45\textwidth]{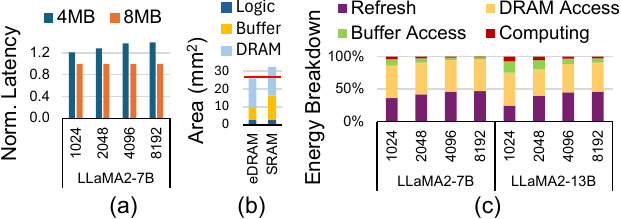}
    \caption{(a) Normalized latency of edge systems with 4MB vs. 8MB SRAM across models and sequence lengths. (b) Area breakdown of the edge systems with 8MB eDRAM and 8MB SRAM. Red line is the area budget. (c) Energy breakdown of the edge system integrating eDRAM. The decoding lengths are shown with a prefilling length of 512. An 8MB eDRAM is used to store KV cache for a subset of layers during decoding. The reported DRAM energy accounts for both model weight access and KV cache offloading from eDRAM.
    }
    \label{fig:motivation}
\end{figure}



\section{Why Use eDRAM for LLMs on Edge Devices?}
\subsection{Benefits and Challenges of Expanding On-Chip Memory}
\label{sec:motivation:edram}
As shown by earlier research~\cite{zhang2024llmcompass, zhao2024alisa, yu2024cambricon}, the speed of serving LLMs is significantly constrained by the bandwidth of off-chip memory. In particular, accessing the KV cache poses the most critical bottleneck during the decoding stage of LLMs~\cite{zhao2024alisa, Zhang2023H2OHO, lee2024infinigen}. A straightforward approach to minimize off-chip memory usage is to expand the on-chip SRAM size, which decreases expensive off-chip memory accesses and boosts system performance~\cite{tu2018rana, chen2014dadiannao}. To illustrate this, we evaluate the latencies of two edge computational systems with 4MB and 8MB of SRAM executing the LLaMA2-7B models across different sequence lengths. Tests are conducted on a simulated platform with a $32\times 32$ systolic array for 8-bit MAC operations, and 16GB DRAM with 64GB/s bandwidth, reflective
of an edge tensor processing unit (TPU) similar to
the Google Coral edge device~\cite{suryavansh2020google}. 
As shown in Figure~\ref{fig:motivation} (a), doubling the SRAM size leads to an average of $1.27\times$ speedup. However, expanding the SRAM capacity from 4MB to 8MB in the evaluation platform increases the power consumption and chip area by $29\%$ and $26\%$, respectively. Given the limited area and power budget for edge environment, increasing SRAM size reduces the resource available for other critical components, leading to a suboptimal system performance~\cite{kung1986memory, wang2022exploration, chen2019deep}. Therefore, we have the following observation:



\researchquestion{
\textbf{Obs. 1}: Larger on-chip memory alleviates the KV caching bottleneck in LLM, but brings area and power penalties in edge devices using SRAM as the on-chip storage. }

\subsection{Pros and Cons of Integrating eDRAM}

To increase the on-chip memory size without increasing area, one approach is to replace SRAM with eDRAM. eDRAM not only provides more than twice the capacity under the same area of SRAM, but also consumes lower access and leakage energy according to Table~\ref{tab:sram_edram}. As shown in Figure~\ref{fig:motivation} (b), the evaluation system with 8MB eDRAM takes less area than the system with 8MB SRAM, leading to lower LLM serving latency within a smaller chip size. Extensive research~\cite{xiao2024high, agrawal2014mosaic, cho2014edram} and commercial products~\cite{giterman20201, kingstonEmbedded, wendel2010power7, fluhr201412, eetimes2013edram} demonstrate the feasibility of integrating eDRAM as the primary on-chip storage medium. However, its potential to benefit LLM serving in edge devices has not been explored.
\begin{figure}[t]
\centering
\begin{minipage}{0.6\columnwidth}
    \centering
    \includegraphics[width=0.88\linewidth]{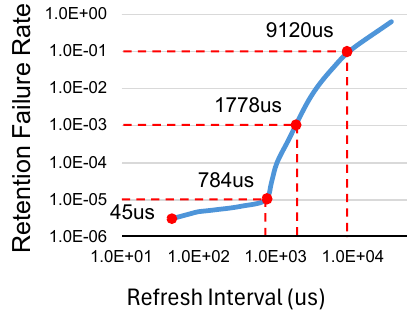}
    \caption{65nm eDRAM retention failure distribution at $105^{\circ}C$~\cite{edram_distribute}.}
    \label{fig:edram_error_distribution}
\end{minipage}%
\hfill
\begin{minipage}{0.38\columnwidth}
    \centering
    \includegraphics[width=\linewidth]{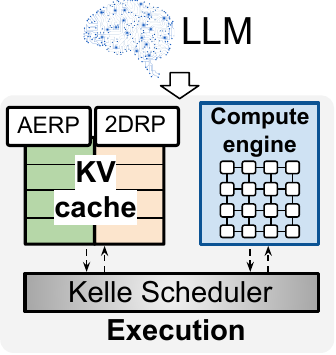}
    \caption{Overview of the Kelle accelerator. }
    \label{fig:overview}
\end{minipage}
\end{figure}
Although eDRAM offers several advantages, prior studies~\cite{tu2018rana, zhang2024camel} have shown that refresh operations can become a significant bottleneck in overall system energy consumption. Moreover, when eDRAM is used to store data with longer lifespans, infrequent refreshes can elevate the risk of readout errors, as illustrated in Figure \ref{fig:edram_error_distribution}. The retention failure rates are represented as the percentage of bits with retention errors, as the refresh interval varies. To illustrate this issue, we use an 8MB eDRAM to replace the 4MB SRAM in the system described in Section~\ref{sec:motivation:edram}. The eDRAM refresh interval is set to $45\mu s$ to ensure no data corruption. We evaluate the energy consumption of the eDRAM system across different models and sequence lengths. As shown in Figure~\ref{fig:motivation} (c), without optimization, eDRAM refresh operations take up to $46\%$ of the total energy consumption, leading to $1.7\times$ more energy consumption on average.



\researchquestion{\textbf{Obs. 2}: Under the same chip area, eDRAM can bring latency benefits over SRAM for LLM serving in edge devices. However, to fully leverage its power advantages, eDRAM refresh operations must be greatly minimized.
}

\subsection{Kelle: Co-design KV Caching and eDRAM}
\label{sec:motivate:codesign}
To minimize eDRAM energy consumption, three effective strategies are: reducing data refresh frequency, decreasing stored data size, and decreasing data lifetime. To enable eDRAM for enhanced LLM serving performance in edge devices, we propose~\textit{Kelle}, a hardware and algorithm co-design solution for minimized eDRAM refresh energy and efficient KV cache management. 
\subsubsection{eDRAM Refresh Control} Reducing data refresh frequency may raise the risk of retention failures, causing data corruption. This leads to a key question: \textit{To what extent can LLMs tolerate data corruption in the KV cache without compromising accuracy?} Led by this question, we co-design the eDRAM memory layout and controller with~\textit{two-dimensional adaptive refresh policy} (2DRP), which sets fine-grained dynamic refresh intervals described in Section~\ref{sec:2drp}.

\subsubsection{KV Cache Eviction} A smaller KV cache can significantly reduce the data storage demands on eDRAM, resulting in lower refresh energy consumption and improved system performance. Previous studies have observed that evicting unimportant tokens does not compromise the generation quality. However, to identify the unimportant tokens, previous works either require profiling of the sequences~\cite{Ge2023ModelTY, Liu2023ScissorhandsET}, or extra computation~\cite{Zhang2023H2OHO, xiao2024efficientstreaminglanguagemodels}. To manage the KV cache efficiently, we propose a novel \textit{systolic evictor} architecture to accelerate the operation of~\textit{attention-based eviction and recomputation policy} (AERP), as described in Section~\ref{sec:aep}. 

\subsubsection{KV Vector Recomputation} As the sequence length grows, the benefit of KV caching diminishes at a certain threshold since the time for accessing off-chip memory might outweigh that for recomputing partial KV tensors. Recomputation aligns well with the strength of eDRAM to store transient data, as shown in Section~\ref{sec:aep}. However, the balance between recomputation and storage requires careful scheduling considering the hardware features. We propose the~\textit{Kelle Scheduler} to reduce the KV vector data lifetime via designing the computational pattern, which is depicted in Section~\ref{sec:Kelle-compiler}.

\section{Kelle Algorithm}
\label{sec:Kelle-algorithm}
In this section, we present the efficient algorithms used within the Kelle framework, with an overview provided in Figure~\ref{fig:overview}. 
During execution, Kelle utilizes~\textit{attention-based eviction and recomputation policy} (AERP) and the~\textit{two-dimensional adaptive refresh policy} (2DRP) to manage eDRAM operation, as described in Section~\ref{sec:aep} and Section~\ref{sec:2drp}.

\subsection{Attention-based Eviction and Recomputation Policy}
\label{sec:aep}

We begin by discussing the eviction policy when the eDRAM capacity is reached during the decoding stage. 

\subsubsection{Eviction policy}

For a KV cache with limited capacity, capable of holding up to \( N' \) tokens, during the decoding stage, the arrival of the (N'+1)-th token requires the eviction of the KV vectors \( [k^{h}_{n}, v^{h}_{n}] \) from one of the tokens \( n \) (where \( 1 \leq n \leq N' \)). The KV vectors of the \(h\)-th head and \(n\)-th token to be evicted are selected based on their importance \(s^{h}_{n}\), which is computed by summing the attention scores (Equation~\ref{eqn:kv-computation1}) with all other tokens in KV cache, shown as:
\begin{equation}
\label{eqn:kv-computation3}
s^{h}_{n} = \sum_{1\leq i\leq n} A^{h}_{n,i}
\end{equation}
An example of the eviction process is illustrated in Figure~\ref{fig:kv_cache_policy}. Assume the KV cache has a budget to store a total of \( N' = 4 \) vectors. We consider a case with three attention heads. For clarity, we only depict the computation for the first head and omit the head notation. When \( [k_{5}, v_{5}] \) arrives, the importance scores are first computed using Equation~\ref{eqn:kv-computation3}, as shown in Figure~\ref{fig:kv_cache_policy} (a). The KV vectors of the token with the smallest importance score (third token) are then evicted, as depicted in Figure~\ref{fig:kv_cache_policy} (b). By leveraging the fact that the computation of \( y_{N} \) is unaffected by the relative order of the KV vectors, Equation~\ref{eqn:kv-computation1} and Equation~\ref{eqn:kv-computation2} can be computed by reading the KV vectors from the cache in sequence, without concern for their original token indices. It is important to note that the importance score \(s_{n}^{h}\) of the same token $n$ might vary across different attention heads. As a result, the eviction pattern of KV vectors will differ across these heads h. 

For the pre-filling stage with a context token length of $N_{cxt}$, all the context tokens are processed in parallel. For each head within each layer, importance scores of Nth token are calculated as $s^{h}_{N} = \sum_{1\leq n\leq N_{cxt}} A^{h}_{n,N}$. The tokens with top $N'$ highest $s^{h}_{n}$ will be retained in the KV cache for decoding operations. 

In addition to the tokens with the highest $s^{h}_{n}$ scores, the initial tokens and most recent tokens are also retained due to their proven impact on model performance, as demonstrated by prior work~\cite{xiao2024efficientstreaminglanguagemodels, Zhang2023H2OHO} and supported by our experiments.

\subsubsection{Recomputation policy}
\label{sec:algo:recomp}

\begin{figure}
    \centering
    \includegraphics[width=0.95\columnwidth]{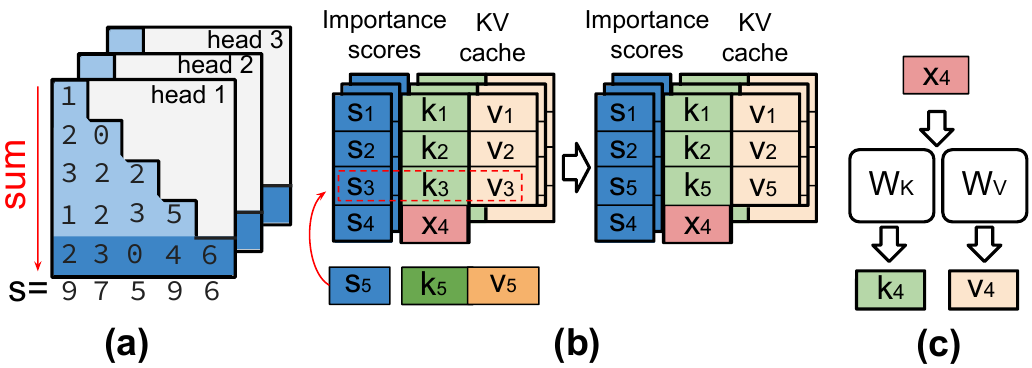}
    \caption{(a) Computation of the importance scores for each of three heads. (b) The KV vectors of the token with the lowest score is replaced with the new KV vectors. Input vector $x_{4}$ is stored because the fourth token is important among two out of three heads. Storing $x_{4}$ frees up an eDRAM entry, thereby reducing the eDRAM refresh cost. (c) Recompute KV vectors of the fourth token for saving eDRAM storage.}
    \label{fig:kv_cache_policy}
\end{figure}

As discussed in Section~\ref{sec:edram_intro}, eDRAM is well-suited for storing transient data. Although the eviction policy reduces the number of KV vectors that need to be retained during model execution, storing KV vectors with long lifetime in eDRAM still incurs a serious cost due to the required refresh operations. To mitigate the refresh cost, we can further apply the recomputation technique. Specifically, for a subset of tokens $N_{recomp}$ in the KV cache, their KV vectors will be recomputed using the corresponding input vector \(x_{N}\), which serve as inputs to \(W_{Q}, W_{K},\) and \(W_{V}\), as depicted in Figure~\ref{fig:llm-intro2} (b). By leveraging recomputation, the storage requirement can be reduced from maintaining two vectors (K and V) to holding just a single vector (input x). This approach allows for K and V to be recomputed as needed, effectively mitigating the long data lifetime of KV vectors. 


During decoding stage execution, the KV vectors \(k^{h}_{N}\) and \(v^{h}_{N}\) for all heads $h\in H$ are first recomputed by multiplying the input vector \(x_{N}\) with \(W_{K}\) and \(W_{V}\) (Figure~\ref{fig:kv_cache_policy} (c)), which are then used for the decoding process. To achieve savings in KV cache storage through recomputation, the storage cost of the input vector \(x_{N}\), which is \(1 \times C\), must be smaller than that of the recomputed KV vectors. To satisfy this, the KV vectors for a token \(N\) are recomputed using \(x_{N}\) if they would otherwise be retained in at least \(\theta > 50\%\) of the heads without recomputation, where \(\theta\) represents the \textbf{popularity} of the token. This approach is justified because the storage cost associated with the KV vectors, calculated as \(2 \times \frac{C}{H} \times \theta H\), would exceed the size of \(x_{N}\) (i.e., \(C\)).
As shown in Figure~\ref{fig:kv_cache_policy} (b), the fourth token is deemed popular in two of three heads, so the input vector \(x_{4}\) is retained, avoiding the storage of KV vectors.

In addition to storage savings, the recomputed KV vectors will be transient, as they are only used for a short duration to compute Equation~\ref{eqn:kv-computation1}, which further capitalizes on the advantages of eDRAM. Moreover, the additional cost of recomputation will be minimal due to the systolic array architecture for the computational engine, discussed in Section~\ref{sec:systolic_array}.

For the pre-filling stage, the importance scores \(s_{n}^{h}\) for each token \(n\) in head \(h\) are calculated first. Next, for each head \(h\), KV vectors are evicted based on the importance scores of the corresponding tokens. Among tokens with high importance scores, those whose KV vectors are retained in at least $50\%$ of the heads (i.e., popular tokens) have their input vectors \(x_{n}\) stored; otherwise, the KV vectors are stored. During decoding, each new token’s storage format is dynamically determined by computing the popularity $\theta$. Figure~\ref{fig:rf_ctrl} (a) summarizes the overall AERP scheme. Although token popularity can vary during the decoding process, empirical evidence shows limited fluctuation, namely tokens important to over $50\%$ of heads rarely decrease in importance. Therefore, in Kelle, once a token is stored with its input vectors, its storage format remains fixed throughout decoding unless it gets evicted.


\subsection{Two-Dimensional Adaptive Refresh Policy}
\label{sec:2drp}


\begin{figure}
    \centering
    \includegraphics[width=0.45\textwidth]{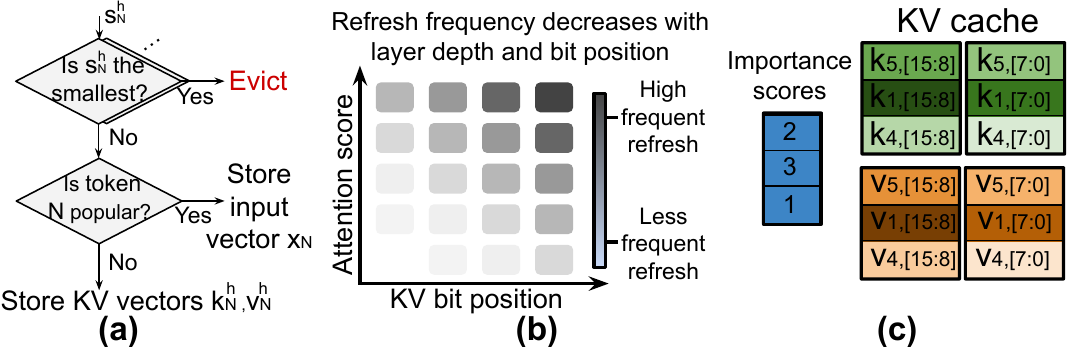}
    \caption{(a) Summary of AERP, only one head h is shown for simplicity. (b) 2D-adaptive refresh policy. (c) As an example on 2DRP. \( k_5[15:8] \) denotes the eights to fifteenth bits of the key vector for the fifth token. Darker colors mean bits refreshed more frequently, resulting in a lower retention error rate.}
    \label{fig:rf_ctrl}
\end{figure}

\begin{figure}
    \centering    \includegraphics[width=0.48\textwidth]{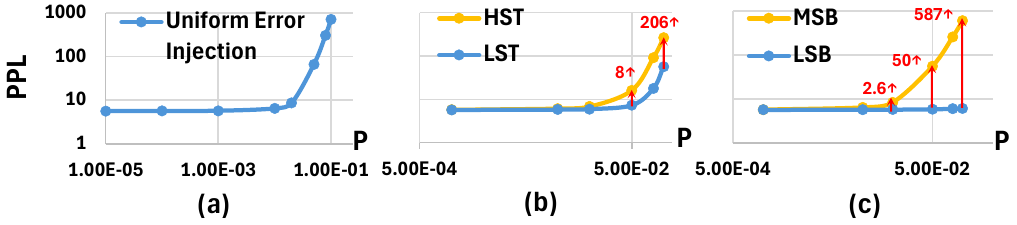}
    \caption{(a) PPLs with bit-flip error rates P. (b) LLM accuracy under varying bit-flip error rates when applying the bit flipping solely on (a) HST vs. LST and (b) MSB vs. LSB, where \( P \) denotes the error rate. A lower PPL reflects better performance, with the red numbers representing the gap between the PPL values.}
    \label{fig:bf_sens}
\end{figure}

To explore the tolerance of LLMs to data corruption in the KV cache without compromising accuracy, we simulate retention failures by introducing bit flip errors across the eDRAM memory cells. Specifically, we assess the impact on the perplexity (PPL) of LLaMA2-7B models using the Wikitext-2~\cite{wikitext} dataset. Lower PPL indicates better performance. During execution, the bit flip errors are introduced in the KV cache with a uniform probability. 
The results, presented in Figure~\ref{fig:bf_sens} (a), reveal that for error rates below $10^{-3}$, the increase in PPL remains minimal, staying under 0.1. However, as the bit flip error continues to rise, PPL increases significantly. This suggests that LLMs can tolerate a certain degree of KV cache errors. A natural follow-up question arises:~\textit{Is it possible to develop a finer-grained refresh policy that could support even lower refresh frequencies while maintaining accuracy?}

In Section~\ref{sec:aep}, tokens are evicted based on their importance scores, as defined in Equation~\ref{eqn:kv-computation3}. We hypothesize that a similar approach could be applied to the eDRAM refresh policy, assigning lower refresh frequencies to KV vectors or input vectors of less important tokens and higher frequency to those of more important tokens. 
To test this hypothesis, we implement an adaptive refresh policy and repeated the experiment. Tokens were divided into two groups based on their importance scores, referred to as the high score token (HST) group and the low score token (LST) group for simplicity.
A probability $p$ of bit retention failure (bit flip error) in the KV vectors was applied separately to the KV vectors of the corresponding tokens in HST and LST groups. 
The results in Figure~\ref{fig:bf_sens} (b) show that LLM performance degrades more when bit retention failures affect the HST group than the LST group, indicating that tokens in the HST group require higher refresh frequency, thus supporting our hypothesis.

Additionally, it is reasonable to hypothesize that the less significant bits (LSBs) are less vulnerable to retention failure errors than the more significant bits (MSBs), as bit flip error on LSBs causes smaller changes in value. 
For each value in a KV vector, we introduce bit retention errors to either the MSBs (bits 15-8) or the LSBs (bits 7-0). The results, shown in Figure~\ref{fig:bf_sens} (c), reveal that the MSBs are more sensitive to retention errors than the LSBs under the same bit flip error rate, further supporting our hypothesis.

Based on observations above, we propose an adaptive refresh control strategy called the~\textit{two-dimensional adaptive refresh policy} (2DRP), as shown in Figure~\ref{fig:rf_ctrl} (b). This strategy adjusts the refresh frequency of each eDRAM cell based on both the bit position within each value in KV vectors or input vectors and the importance score of each token. An example of 2DRP is shown in Figure~\ref{fig:rf_ctrl} (c), where the KV cache holds up to \( N' = 3 \) tokens. The refresh frequency increases with both the token importance and the significance of bit positions. During execution, the importance scores of KV and input vectors are dynamically calculated, and a refresh frequency is assigned accordingly based on these scores and bit positions.

\section{Kelle Edge Accelerator}
\label{sec:Kelle-hardware}

\begin{figure}
    \centering
    \includegraphics[width=0.45\textwidth]{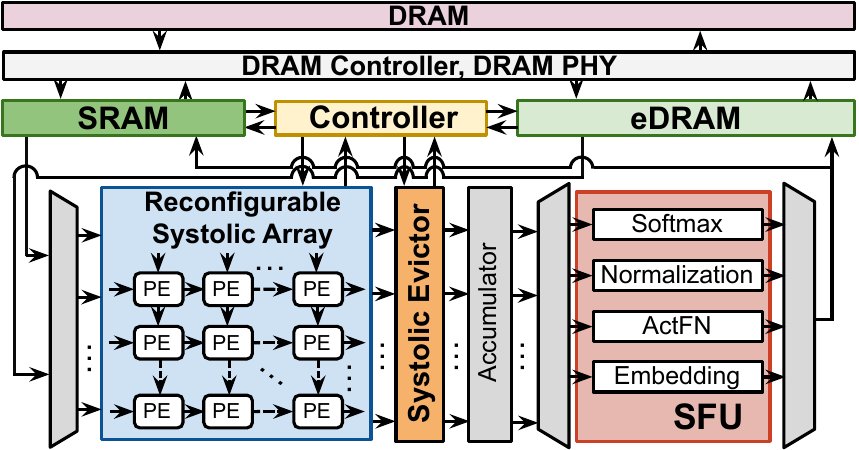}
    \caption{An overview of Kelle hardware accelerator.}
    \label{fig:hw_overview}
\end{figure}
Figure~\ref{fig:hw_overview} provides an overview of the Kelle accelerator. It incorporates a hybrid eDRAM-SRAM memory subsystem, a reconfigurable systolic array (RSA), and specialized function units (SFUs). 
The weights are quantized to 8 bits, and activations and KV vectors are maintained in 16 bits, with weights stored in SRAM, while activations and the KV vectors are held in eDRAM. During operation, systolic evictor accumulates attention scores and eDRAM controller handles KV vector eviction and recomputation while dynamically adjusting the refresh frequency, as discussed in Sections \ref{sec:aep} and \ref{sec:2drp}. Each processing element (PE) in RSA performs 8-bit multiply-accumulate (MAC) operations. 

SFU handles non-linear operations, including activation functions, softmax, normalization, and positional embeddings. As prior research has shown \cite{Wang2020SpAttenES, hyft, dass2022vitalityunifyinglowranksparse,qin2025picachu}, the energy consumption of non-linear operations increases with input sequence length. Among these operations, softmax consumes significant resources. We employ online max calculation from Softermax \cite{stevens2021softermaxhardwaresoftwarecodesignefficient} to minimize memory access.
For other non-linear operations, we follow the computation flow and use lookup tables (LUTs) to perform the calculation.



\subsection{Memory Subsystem}
\label{sec:memory}

\begin{figure}
    \centering
    \includegraphics[width=0.43\textwidth]{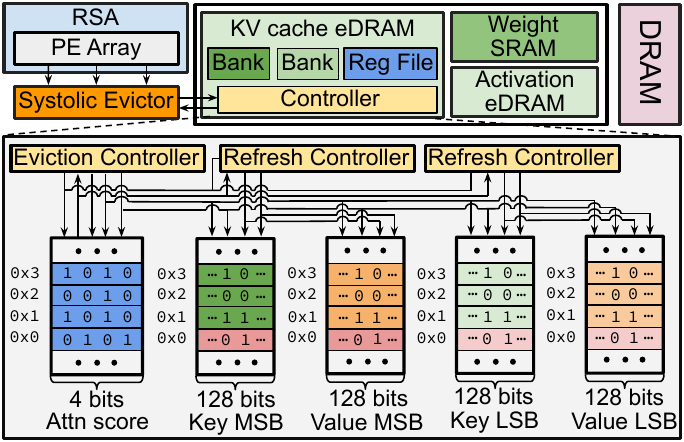}
    \caption{Kelle memory subsystem. Input vectors of certain tokens are stored in the KV cache, represented by red rows.}
    \label{fig:memory}
\end{figure}

Figure~\ref{fig:memory} illustrates the memory subsystem of the Kelle accelerator. In this design, a 2MB SRAM stores the weights, while activations and KV vectors are held in a 256KB~\textit{activation eDRAM} and 4MB~\textit{KV cache eDRAM}, respectively. Kelle accelerator implements 2DRP by dividing KV vectors into four groups based on importance scores and bit positions and applies the refresh frequency accordingly. Specifically, the MSBs (bits 15-8) of the KV vectors for tokens in the HST group are refreshed at the highest, while LSBs (bits 7-0) of the KV vectors for tokens in the LST group are refreshed at the lowest frequency. To support the AERP, for certain tokens, input vectors are stored in the KV cache eDRAM instead of KV vectors. These input vectors are then divided into four groups and controlled in the same manner as KV vectors, organized by importance scores and bitwidth. For simplicity, we use KV vectors to describe the memory subsystem design, without referencing input vectors.

To execute 2DRP during LLM inference, each element of the KV vectors is split bitwise and stored across different eDRAM banks. Specifically, for KV vectors, the MSBs and LSBs are stored in separate KV cache eDRAM banks, referred to as~\textit{MSB banks} and~\textit{LSB banks}, highlighted by the darker and lighter colors in Figure~\ref{fig:memory}. The importance score of each token is computed dynamically using Equation~\ref{eqn:kv-computation3} under 4-bit precision and stored in a register file, with each entry corresponding to a KV vector spanning four banks. KV vectors corresponding to the same token share the same address across different eDRAM banks. The system features a single eviction controller that manages AERP across all four banks, along with two refresh controllers responsible for executing 2DRP separately over MSB and LSB banks, respectively.

In each MSB and LSB bank, the tokens are further divided into two groups according to their attention scores, with a counter in the refresh controller monitoring each group's refresh interval. The controller iterates through the eDRAM entries, identifying the group of each token by reading its attention score from the register file. When the refresh interval for a specific group expires, the controller triggers a~\textit{refresh} signal. The corresponding KV vector addresses for the tokens in that group are then computed, and the KV vectors are read out and written back as part of the refresh procedure. The refresh operation is triggered when the KV vectors are not used by the model, so the refresh latency can be hidden.
When the KV cache reaches capacity and a new token arrives, the eviction controller receives the evict token index from the systolic evictor and replaces that token with the new token.

To fully utilize the $32\times32$ RSA by feeding data in parallel and avoiding bank conflicts, the Kelle KV cache is divided into 32 banks. Specifically, 8 banks are assigned to each of the Key MSB, Key LSB, Value MSB, and Value LSB groups. With this design and pipelined cache read, Kelle eDRAM provides sufficient bandwidth to make full use of the RSA without bank conflicts. Additionally, other eDRAM access operations such as token read and token eviction operate independently, this effectively mitigates bank conflicts.

During the LLM execution, the RSA I/O controller efficiently reconstructs the data from different banks for computation with minimal overhead. Additionally, the Kelle accelerator stores the KV vectors from a subset of LLM layers in eDRAM, with the number of layers determined by the specific LLM size and text length. eDRAM greatly minimizes off-chip memory access overhead.

\subsection{Reconfigurable Systolic Array}
\label{sec:systolic_array}

\begin{figure}
    \centering
    \includegraphics[width=0.95\columnwidth]{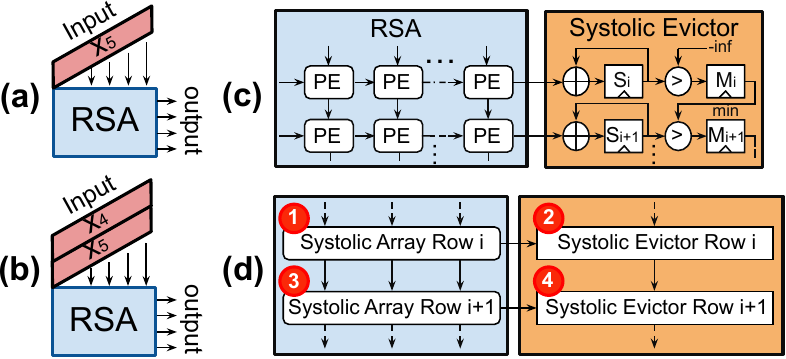}
    \caption{(a) and (b) show the impact of recomputation on RSA operation. (c) The integration of RSA and systolic evictor. (d) The execution order of systolic array and systolic evictor. Numbers in red circles denote the orders.}
    \label{fig:pe}
\end{figure}

The systolic array core consists of a $32\times 32$ two-dimensional array that processes inputs in a staggered manner, sending the computed partial sums to the accumulator and SFUs. It utilizes a weight-stationary data flow, as shown in Figure~\ref{fig:pe} (a).
We employ a reconfigurable strategy similar to that in FAST~\cite{Zhang2021FASTDT} to perform in-place transposed matrix multiplication. 

Importantly, the recomputation in Section~\ref{sec:aep} introduces minimal overhead in the LLM decoding stage. Leveraging the strength of systolic arrays for matrix operations, the recomputed token vectors can be combined with the current token’s input vector to create an input matrix efficiently. Using the same notation as the example in Figure~\ref{fig:kv_cache_policy} (b), Figure~\ref{fig:pe} (a) shows the input vector $x_{5}$ of the current token being sent to RSA for KV vector computation. To recompute KV vectors for the fourth token, $x_{4}$ and $x_{5}$ can be combined into a matrix, causing minimal latency and energy growth in Figure~\ref{fig:pe}(b).

\subsection{Systolic Evictor}
\label{sec:evictor}
\begin{figure*}
    \centering
    \includegraphics[width=0.9\textwidth]{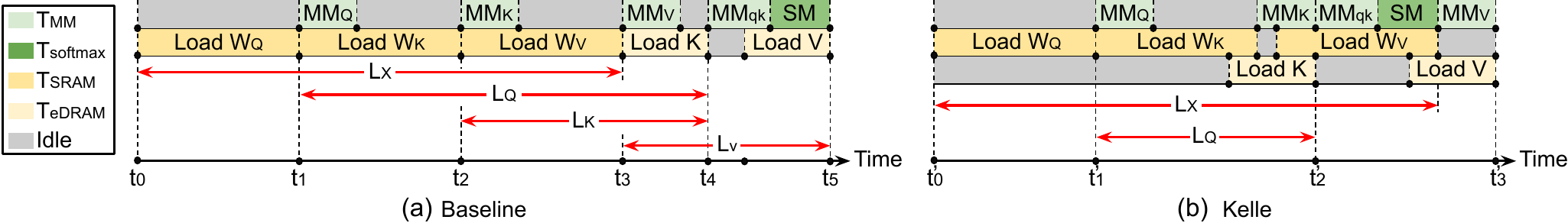}
    \caption{(a) and (b) show the computation patterns and eDRAM data lifetime of the SA block in baseline and Kelle scheduler. SM denotes softmax operation.}
    \label{fig:comp_pattern}
\end{figure*}
The token eviction process in the AERP algorithm includes computing the attention score as described in Equation~\ref{eqn:kv-computation1}, updating the importance score based on Equation~\ref{eqn:kv-computation3}, identifying the token with the lowest importance score, and performing the KV cache update.

To efficiently implement the eviction algorithm, we propose a systolic evictor (SE) which operates in a systolic style and is integrated with the RSA to search the minimum importance score on-the-fly. The importance score is calculated by summing the $QK^T$ results in Equation~\ref{eqn:kv-computation1} without passing through the softmax. This integration ensures the token with the minimum importance score is found as soon as the new token's attention score is calculated from the RSA. After finding the index of the token with minimum importance score, the SE sends the index to the eviction controller in the eDRAM controller to evict the corresponding token. Figure~\ref{fig:pe} (c) illustrates the design of SE and its integration in the RSA. The SE comprises a column of registers, denoted as~\textit{S} in Figure~\ref{fig:pe} (c), to preload the importance scores of previous tokens, and a register chain, denoted as~\textit{M}, periodically propagates the minimum importance score (min) from top to bottom. Figure~\ref{fig:pe} (d) illustrates execution order of the RSA and SE. In one cycle, the attention score is calculated from the $i$-th row of RSA, and then the $i$-th row of SE updates the importance score and the minimum importance score index, marked as Step 1 and Step 2, respectively. In the next cycle, the same operations are executed in the next row of RSA and SE, marked as Step 3 and Step 4, respectively. The systolic evictor avoids the extra LLM execution latency from the minimum search.

\section{Kelle Scheduler}
\label{sec:Kelle-compiler}
To further minimize eDRAM refresh energy, we introduce a novel computation pattern that shortens data lifetime and accelerates LLM inference, all without compromising accuracy. 

To begin, we perform a numerical analysis of the data lifetime associated with the self-attention (SA) architecture in the LLM decoding phase. As illustrated in Section~\ref{sec:bg:workflow}, the computation in SA firstly involves matrix multiplication between the input $X$ with weight matrices $W_{Q}$, $W_{K}$, and $W_{V}$ producing the output $Q, K$ and $V$. The processes are denoted as $MM_Q$, $MM_{K}$, and $MM_{V}$, respectively. Subsequently, $Q$ and $K$ undergo multiplication, followed by a softmax operation to compute the attention score $A$, which are referred to as $MM_{qk}$ and $SM$, respectively. Ultimately, $A$ is multiplied by the weight matrix $W_{O}$ to produce the SA output, labeled as $MM_O$. The latency ($T_{MM}$) of matrix multiplications is estimated as follows:
\begin{equation}
\label{eqn:tmm}
    \small
    T_{MM} = \frac{N_{MM}}{TOP_{RSA}}
\end{equation}
where $N_{MM}$ represents the number of MAC operations required by the matrix multiplication. $TOP_{RSA}$ denotes the throughput of the RSA, as described in Section~\ref{sec:systolic_array}.
The latency associated with eDRAM access operations for KV vectors, represented as \( T_{eDRAM} \), is modeled as follows:
\begin{equation} 
\label{eqn:tedram}
\small
T_{eDRAM} = \frac{S_{KV}}{B_{eDRAM}}
\end{equation}
where $S_{KV}$ denotes the size of the KV vectors in bytes. $B_{eDRAM}$ denotes the bandwidth of the eDRAM. Similarly, the latency associated with the weight SRAM access operations, denoted as $T_{SRAM}$, is modeled as follows:
\begin{equation} 
\label{eqn:tsram}
\small
T_{SRAM} = \frac{S_W}{B_{SRAM}}
\end{equation}
where $S_W$ denotes the size of weights in bytes. $B_{SRAM}$ denotes the bandwidth of SRAM. 
\begin{table*}[htbp]
 \caption{Accuracy performance of each method. FP16 denotes the LLM accuracy under FP16 without KV cache reduction. 
 }
\begin{center}
\resizebox{2\columnwidth}{!}{%
\begin{tabular}{c|ccccc|ccccc|cccc|cc|cc|ccc|ccc}\hline
Model & \multicolumn{5}{c|}{LLaMA2-7B} & \multicolumn{5}{c|}{LLaMA2-13B} & \multicolumn{4}{c|}{LLaMA3.2-3B} & \multicolumn{2}{c|}{LLaMA3-8B} & \multicolumn{2}{c|}{Mistral-7B}  & \multicolumn{3}{c|}{QWEN2-7B} & \multicolumn{3}{c}{OPT-6.7B}\\  
     Method & FP16  & SL   & H2O   & QR & \textbf{Kelle}  & FP16  & SL  & H2O  & QR    & \textbf{Kelle}  & FP16 & SL & H2O  & \textbf{Kelle} &  FP16 & \textbf{Kelle} & FP16 & \textbf{Kelle} & FP16 & H2O & \textbf{Kelle}  & FP16 & H2O & \textbf{Kelle}\\ \hline  
     WK2 ($\downarrow$)& 5.47 & 6.89 & 5.70 & 5.73  & 5.74  & 4.88 & 6.21 & 5.44  & 5.83 & 5.62 & 6.32 & 7.93 & 6.57  & 6.65  & 6.14 &  6.59  & 5.25 &  5.86 &  5.32 & 6.11 & 6.23 & 12.3 & 13.8 & 14.6\\ 
     PG19 ($\downarrow$) & 10.51 & NA & 12.34 & 11.77 & 12.59 & 8.75 & NA & 9.84 & 9.15  & 9.80  & 8.95 & NA & 9.84  & 9.66 & 11.82  & 12.97 & 10.42 &  13.30 & 9.3 & 11.2 & 11.4 & 17.4 & 20.1 & 19.8\\
     A-c ($\uparrow$) & 46.33 & 38.40 & 46.10 & 45.80  & 45.93 & 49.06 & 46.08 & 48.35 & 47.59  & 48.83 & 50.34 & 46.59 & 50.08  & 50.03  & 53.16  & 51.89 & 55.03  & 54.98 & 57.6 & 56.1 & 55.8 & 45.2 & 44.1 & 44.8\\
     
     A-e ($\uparrow$) & 74.62 & 52.99 & 73.02 & 72.89  & 72.78 & 77.48 & 58.25 & 76.71 & 76.45  & 76.33  & 76.98 & 70.45 & 76.31  & 76.45 & 77.69 & 75.81 & 80.22  & 78.90 & 70.5 & 67.8 & 68.2 & 60.1 & 58.8 & 58.6\\
     
     PQ ($\uparrow$) & 79.11 & 75.14 & 78.05 & 77.42  & 77.35 & 80.52 & 78.92 & 79.05 & 78.78  & 78.92 & 79.7 & 75.78 & 79.43  & 78.64 & 80.52  & 77.62 & 82.15  & 80.81 & 80.9 & 78.4 & 78.5 & 76.4 & 75.2 & 74.9\\
     
     
     LA ($\uparrow$) & 73.90 & NA & NA & 72.39  & 72.81 & 76.75 & NA & NA & 75.67  & 75.98  & 73.58 & NA & NA  & 71.26 & 75.72  & 73.38 & 75.49  & 75.54  & 68.4 & NA & 67.1 & 61.5 & NA & 59.2 \\  
     
     TQ ($\uparrow$) & 48.95 & NA & 47.53 & 47.56  & 47.40 & 58.73 & NA & 57.46 & 56.86  & 57.55 & 59.84 & NA & 58.62  & 58.70  & 61.53  & 58.79  & 61.57  & 59.96 & 63.5 & 61.1 & 60.7 & 43.2 & 40.3 & 40.6\\ 
     QP ($\uparrow$) & 12.69 & NA & 12.31 & 12.06  & 12.18 & 13.07 & NA & 11.75 & 11.63  & 11.86 & 12.52 & NA & 10.87  & 11.01  & 13.84  & 11.88  & 13.18  & 11.67 & 21.3 & 19.4 & 19.5 & 9.3 & 7.8 & 7.5\\
     \hline
\end{tabular}
}
\label{tal:acc-eval}
\end{center}
\end{table*}
Figures \ref{fig:comp_pattern} (a) show a baseline computation pattern, where the matrix multiplication operations $MM_{Q}$, $MM_{K}$, $MM_{V}$, and $MM_{qk}$ are conducted one after another, which prolongs the data lifetime for the inputs $X$, $Q$, $K$, and $V$. Data lifetime is defined as the interval from when data is computed to when it is utilized by a subsequent operation. For instance, in Figures \ref{fig:comp_pattern} (a), the computation of vector \(Q\) starts at \(t_1\) after the weight matrix \(W_Q\) is accessed from SRAM, and \(Q\) is consumed at \(t_4\) when the multiplication of \(Q\) and \(K\) starts. Between \(t_1\) and \(t_4\), \(W_K\) and \(W_V\) are loaded from SRAM and \(K\) is accessed from eDRAM KV cache. The latencies of accessing \(W_K\) and \(W_V\) are both \(T_{SRAM}\) and the latency of accessing \(K\) is \(T_{eDRAM}\). So the data lifetime of \(Q\) is \(2 \times T_{SRAM}+T_{eDRAM}\). The total data lifetime of all the activations is the sum of the data lifetime of each activation because they are all stored in eDRAM and require refreshing. We omit the computation time \(T_{MM}\) from Equation~\ref{eqn:tmm} due to its negligible magnitude relative to \(T_{SRAM}\) and \(T_{eDRAM}\). This extended data lifetime leads to a higher refresh cost for eDRAM. The total data lifetime $L_{bl}$ of the transient data in baseline schedule is modeled as follows:
\begin{equation}
\small
\begin{split}        
        L_X  & = 3 \times T_{SRAM} , L_Q = 2 \times T_{SRAM} + T_{eDRAM} \\ 
        L_K  & = T_{SRAM} + T_{eDRAM}, L_V = 2T_{eDRAM} \\
        L_{bl} &= L_X + L_Q + L_K + L_V = 6T_{SRAM}+4T_{eDRAM} 
\end{split}
\end{equation}
where $L_X$, $L_Q$, $L_K$ and $L_V$ denotes the data lifetime of $X$, $Q$, $K$ and $V$, respectively. \(T_{SRAM}\) and \(T_{eDRAM}\) are defined in Equation~\ref{eqn:tsram} and Equation~\ref{eqn:tedram}. 
In contrast, the computation pattern used by the Kelle is illustrated in Figures \ref{fig:comp_pattern} (b). Thanks to the integration of separate on-chip memory, the memory access for weights and KV vectors is parallelized. This arrangement reduces the data lifetime of activations, which can be estimated as follows:
\begin{equation}
\small
\begin{split}  
        L_X  & = 3 \times T_{SRAM} , L_Q = T_{SRAM} + T_{eDRAM} \\ 
        L_{Kelle} & = L_X + L_Q   = 4T_{SRAM}+T_{eDRAM} 
\end{split}
\end{equation}
The key and value vectors are used immediately for their respective computations, eliminating the need for long-term storage; therefore, their data lifetimes can be considered negligible. Compared to the baseline scheme, the Kelle scheduler significantly reduces the data lifetime of transient data in eDRAM, resulting in decreased refresh energy consumption and enhanced system performance. 

\section{Accuracy Evaluation}
\label{sec:accuracy-eval}


\subsection{Main Accuracy Result}
\label{sec:accuracy-main-eval}

Kelle is evaluated on various LLMs, including Llama2~\cite{touvron2023llama}, Llama3~\cite{dubey2024llama3herdmodels}, Llama3.2~\cite{dubey2024llama3herdmodels}, Mistral~\cite{jiang2023mistral7b}, QWEN~\cite{yang2024qwen2technicalreport}, and OPT~\cite{zhang2022optopenpretrainedtransformer} with varying model sizes. We evaluate Kelle on the language generation tasks by perplexity of WikiText-2 (WK2)~\cite{wikitext}, and PG19~\cite{rae2019compressive}. WK2 sequences range from hundreds to thousands of tokens.
PG19 sequences range from tens of thousands to millions. We evaluate the PG19 text generation task using the Cold Compress framework~\cite{gpt-fast, cold-compress-2024} by providing the model a book title and a short description and setting the sequence generation length to 8192.
Kelle is also evaluated over different zero-shot tasks, including PIQA (PQ)~\cite{Bisk2019PIQARA}, Lambada (LA)~\cite{Radford2019LanguageMA}, Arc Easy (A-e)~\cite{Clark2018ThinkYH}, Arc Challenge (A-c)~\cite{Clark2018ThinkYH}, TriviaQA (TQ)~\cite{joshi2017triviaqa}, and Qasper (QP)~\cite{dasigi2021dataset}. We use the LM Evaluation Harness~\cite{eval_harness} with default parameters. 

For KV vector eviction, the number of tokens retained in the KV cache is dynamically adjusted based on the dataset during both pre-filling and decoding phases. 
To simulate bit flip error from low eDRAM refresh frequency, bit-level retention failure is introduced with a predefined probability based on the refresh interval. We set the refresh interval as 0.36ms, 5.4ms, 1.44ms, and 7.2ms for the MSBs (bits 15-8) of HST, LSBs (bits 7-0) of HST, MSBs of LST, LSBs of LST, respectively, with an average retention time of 1.05ms. This achieves an averaged retention failure rate at 2e-3.

\begin{table}
   \caption{LLaMA2-7B accuracies over different cache sizes. }
    \centering
    \resizebox{0.85\columnwidth}{!}{%
       \begin{tabular}{c|c|c|c|c|c|c|c}
        \hline
    $N'$ & Full & 512 & 256 & \textbf{128} & 64 & 32 & 16 \\ \hline

    A-c ($\uparrow$)      & 46.02 & 46.02 & 45.92 & 45.93 & 44.63 & 44.20 & 38.52 \\ 
    A-e ($\uparrow$)     & 73.05 & 73.04 & 73.05 & 72.78 & 72.38 & 70.42 & 67.27 \\ 
    PQ ($\uparrow$)      & 78.75 & 78.49 & 77.61 & 77.35 & 75.31 & 74.14 & 71.63 \\ 
    \hline
    \end{tabular}
}
\label{tab:cache_size}
\end{table}


We compare the accuracy performance of the Kelle algorithm with a state-of-the-art quantization framework, QuaRot (QR)~\cite{ashkboos2024quarot}. Additionally, we include StreamLLM~\cite{xiao2024efficientstreaminglanguagemodels} and H2O~\cite{Zhang2023H2OHO}, two recent KV cache eviction techniques, for comparison. 
The model weights are quantized with 8-bit across all the approaches. To keep the same KV cache budget between quantization and KV cache eviction baselines, we configure QuaRot to quantize the KV vectors to 4-bit, while StreamLLM, H2O, and Kelle are left unquantized as 16-bit. For Kelle, we maintain a token storage budget of N'=128 for the PQ, LA, A-e and A-c, N'=512 for WK2, N'=1024 for the TQ and QP, and N'=2048 for PG19. Within the token budget, the most recent token window size is configured as 64 for PQ, LA, A-e, and A-c; 256 for WK2; 512 for TQ and QP; and 1024 for PG-19. 10 initial tokens are also preserved across the datasets. StreamLLM and H2O are set to have the same token storage budget as Kelle. We also compare with the original FP16 models without KV cache eviction, denoted as FP16. As shown in Table~\ref{tal:acc-eval}, Kelle maintains a comparable accuracies as the original full KV cache model and outperforms or achieves a comparable performance as the rest methods, showing the superior accuracy performance of AERP and 2DRP algorithms. 

\subsection{Ablation Study}
\label{sec:ablation-studies}
We adjust the budget size $N'$ for the Llama2-7B model and examine its impact across various tasks. All other settings (e.g., quantization bitwidth, retention failure rate) for Kelle remain the same. From Table~\ref{tab:cache_size}, we observe a consistent decline in accuracy as the budget $N'$ decreases, but still achieves reasonable performance for $N'\geq 128$, comparing with the KV cache without pruning (Full in Table~\ref{tab:cache_size}). 

\begin{table}
    \caption{LLaMA2-7B accuracies of different refresh intervals. }
    \centering
    \resizebox{0.95\columnwidth}{!}{%
       \begin{tabular}{c|cc|cc|cc}
        \hline 
    Uniform $(\mu s)$  & \multicolumn{2}{c|}{$540$}  & \multicolumn{2}{c|}{$1050$}    & \multicolumn{2}{c}{$2062$}     \\ \hline
    HST  $(\mu s)$ & \multicolumn{2}{c|}{$180, 3600$}  & \multicolumn{2}{c|}{$360, 5400$}    & \multicolumn{2}{c}{$720, 9000$} \\ 
    LST  $(\mu s)$ & \multicolumn{2}{c|}{$720, 5400$} & \multicolumn{2}{c|}{$1440, 7200$}    & \multicolumn{2}{c}{$2880, 10800$} \\ \hline
    Accuracy  & Uniform & 2DRP & Uniform & 2DRP & Uniform & 2DRP  \\ 
    \hline
    A-c ($\uparrow$)  & 45.31 & 46.26 & 44.19 & 45.93   & 38.52  &  39.78   \\ 
    A-e ($\uparrow$)       & 72.85 & 73.12 & 70.29 & 72.78   & 66.50  & 67.05 \\ 
    PQ ($\uparrow$)           & 76.83 & 77.44 & 76.43 & 77.35   & 74.97   & 75.21 \\ 
    \hline
    \end{tabular}
}
\label{tab:flip_rate}
\end{table}

\begin{figure*}
    \centering
    \includegraphics[width=0.99\textwidth]{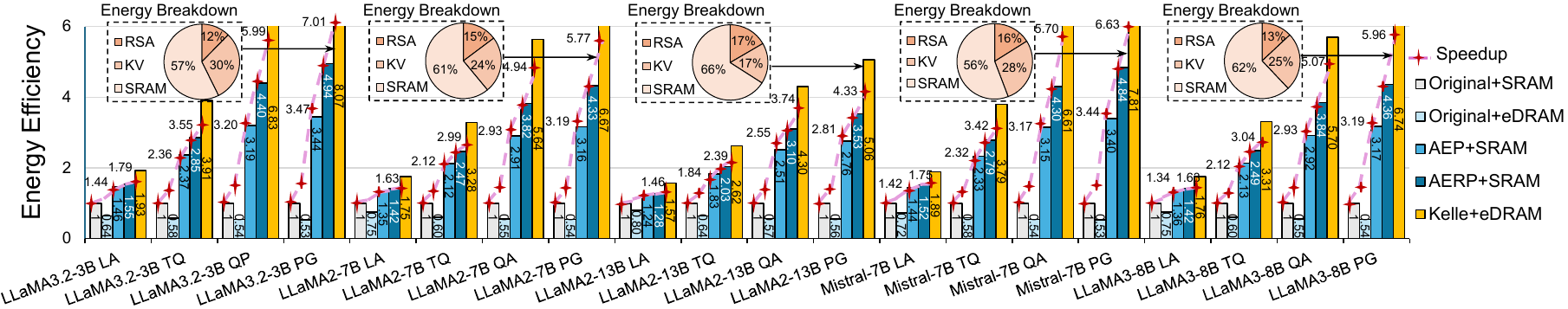}
    \caption{Comparison between Kelle and baseline systems. The performance are evaluated in terms of normalized energy efficiency and speedup. The pie charts show the on-chip energy breakdown of major components within Kelle+eDRAM. The dotted red lines depict the speedup of the corresponding settings. }
    \label{fig:hw_eval1}
\end{figure*}
Next, we examine the impact of 2DRP on LLM accuracy. Specifically, we compare 2DRP with a condition where all eDRAM cells share the same refresh interval, while maintaining the same average retention failure rate as 2DRP. All other conditions remain the same for Kelle. In Table \ref{tab:flip_rate}, we present the accuracy change with varying refresh intervals on different tasks for the Llama2-7B model. In Table~\ref{tab:flip_rate}, Uniform ($\mu$s) denotes the uniform refresh interval applied to eDRAM. The two numbers in the HST row represent the refresh intervals for the MSBs and LSBs of HSTs, same for the LST row. We observe that 2DRP improves accuracy compared to the uniform eDRAM refresh across all the conditions and datasets. 

Since edge deployment often involves human-facing applications, it is important to evaluate the impact of the approximations in memory behavior introduced by 2DRP on text generation qualitative metrics. To evaluate coherence, we run Kelle on the LLaMA2-7B and Mistral-7B models using the CNN/DailyMail~\cite{nallapati2016abstractivetextsummarizationusing} (CNN) summarization dataset and report the ROUGE-1 scores. To evaluate factual correctness, we test Kelle on the TruthfulQA benchmark~\cite{lin2022truthfulqameasuringmodelsmimic} (Truth) and report the multiple-choice, single-answer accuracy. To assess bias tendencies, we use the BBQ benchmark~\cite{parrish2022bbqhandbuiltbiasbenchmark} and report the corresponding bias evaluation scores for both models. The results  in Table~\ref{tab:quality} show that Kelle achieves performance comparable to the FP16 model across all criteria.

Finally, we quantize the Llama2-7B model using the Quarot framework~\cite{ashkboos2024quarot} which adopts Hadamard Transformation to enable low bit LLM quantization. We quantize the model weights to 4-bit, KV vectors, and activations to 8-bit. With quantization, Kelle's system performance is expected to improve further, while the impact on accuracy remains minimal, as shown in Table~\ref{tab:quantize}. This demonstrates Kelle's compatibility with model quantization techniques.

\begin{table}[t]
  \centering
  \begin{minipage}[t]{0.5\columnwidth}
    \centering
    \caption{Kelle Qualitative Metrics}
    \resizebox{\linewidth}{!}{%
      \begin{tabular}{c|c|c|c|c}
        \hline
        Model & Method & CNN & Truth & BBQ \\ \hline
        LLaMA2 & FP16 & 40.58 & 34.28 & 95.21 \\
        7B & Kelle & 38.54 & 33.26 & 93.75 \\ \hline
        Mistral & FP16 & 36.13 & 36.31 & 96.11 \\
        7B & Kelle & 34.61 & 34.89 & 94.83 \\
        \hline
      \end{tabular}
    }
    \label{tab:quality}
  \end{minipage}
  \hfill
  \begin{minipage}[t]{0.45\columnwidth}
    \centering
    \caption{Accuracy of Kelle with Quantization}
    \resizebox{\linewidth}{!}{%
      \begin{tabular}{c|c|c}
        \hline
        Method & Kelle W8A16 & Kelle W4A8 \\ \hline
        WK2 ($\downarrow$) & 5.74 & 6.51 \\
        AC ($\uparrow$) & 45.93 & 44.89 \\
        AE ($\uparrow$) & 72.78 & 69.96 \\
        PQ ($\uparrow$) & 77.35 & 76.70 \\
        \hline
      \end{tabular}
    }
    \label{tab:quantize}
  \end{minipage}
\end{table}

\section{Hardware Evaluation}
\label{sec:hardware-eval}
In this section, we report the hardware evaluation results of Kelle edge accelerator described in Section~\ref{sec:Kelle-hardware}.
The Kelle edge accelerator consists of a 2D \(32 \times 32\) RSA, an SFU, essential interfaces, and a memory controller, all implemented in RTL using SystemVerilog, with frequency set to 1GHz. We report the area and power of Kelle accelerator by synthesizing the components using 45nm NanGate Open Cell Library~\cite{nangate} with Synopsys Design Compiler \cite{Baliga2019Synopsys}. The size of SRAM for weight storage is set to 2MB. The size of eDRAM for KV cache and activation storage is set to  4MB and 256KB respectively. SRAM and eDRAM bandwidths are set to 128GB/s and 256 GB/s, respectively. We utilize Destiny~\cite{destiny} to evaluate the area, power, and timing performance of the eDRAM and SRAM with 65nm tech node at $105^{\circ}C$. The eDRAM retention time distribution aligns with the data shown in Figure~\ref{fig:edram_error_distribution} \cite{edram_distribute, zhang2024camel}. Notably, eDRAM operating at temperatures below \(105^{\circ}C\) exhibits an even longer retention time, further enhancing system performance. We utilize Cacti 7 \cite{cacti} to simulate the performance of a 16GB LPDDR4 DRAM, with 64GB/s bandwidth, similar to the DRAM in
the Google Coral edge device~\cite{suryavansh2020google}. 
With these settings, the total on-chip area is $9.5mm^2$ and the area breakdown of RSA, eDRAM, SRAM, SFU are $23\%$, $33\%$, $37\%$, $7\%$, respectively. The DRAM takes an area of $16mm^2$. The on-chip power is 6.52W and the power breakdown of RSA, eDRAM, SRAM, SFU are $17\%$, $29\%$, $41\%$, $13\%$, respectively. The DRAM power is 11.74W. Kelle accelerator achieves 4.13 INT8 TOPs. The Kelle scheduler described in Section~\ref{sec:Kelle-compiler} further reduces the eDRAM cost.  

We assess the hardware performance of the Kelle accelerator across various LLM architectures over multiples tasks including Lambada (LA)~\cite{Radford2019LanguageMA}, TriviaQA (TQ)~\cite{joshi2017triviaqa}, Qasper (QA)~\cite{dasigi2021dataset}, and PG19~\cite{rae2019compressive}, with the context length set to 128, 512, 1024, and 512, and the decoding length set to 512, 2048, 5120, and 8192, respectively. The batch size is set to 16. The off-chip DRAM access latency and energy are included in all evaluation results. 


\begin{figure}
    \centering
    \includegraphics[width=0.95\columnwidth]{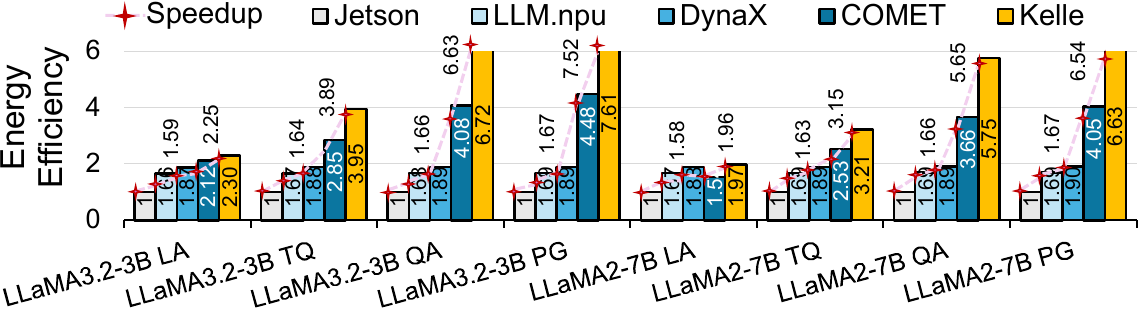}
    \caption{Comparison between LLM accelerators
    }
    \label{fig:hw_eval_accelerators}
\end{figure}

\subsection{End-to-End Performance Evaluation}
\label{sec:end-to-end-accu}
\subsubsection{Evaluation Baseline}
To understand the separate contributions of the Kelle algorithm and eDRAM-based accelerator discussed in Section~\ref{sec:Kelle-hardware}, we compare the Kelle algorithm paired with an eDRAM-based Kelle accelerator, referred to as \textbf{Kelle+eDRAM}, against four baseline solutions.

The first baseline, \textbf{Original+SRAM}, runs the original LLM on a system using SRAM as the primary on-chip storage. The model weights are quantized to 8 bits, activations and KV vectors remain 16 bits and processed using Kelle RSA configured for 8-bit MAC operations. The KV cache remains intact, with no AERP applied. The SRAM-based system is configured to match the total on-chip area of Kelle+eDRAM. We adjust the SRAM and systolic size to achieve balanced compute/memory IO ratio, resulting in a systolic array with $24\times24$ 8-bit PEs, 4MB of on-chip SRAM, 16GB of off-chip DRAM. 
The second baseline, \textbf{Original+eDRAM}, involves running the original LLM on an eDRAM-based Kelle accelerator while keeping the KV cache intact. The models are processed using Kelle RSA configured for 8-bit MAC operations. This baseline removes all algorithmic innovations and evaluates only the performance of the eDRAM-based system.
In the third baseline, \textbf{AEP+SRAM}, we apply the attention-based eviction techniques with the settings described in Section \ref{sec:accuracy-main-eval} for KV cache pruning, implemented on the same SRAM-based system of Original+SRAM. 
The goal is to evaluate the impact of the cache eviction algorithm on the SRAM-based system. Note that this baseline does not involve any recomputation.
The fourth baseline, \textbf{AERP+SRAM}, runs the AERP algorithm on the SRAM-based Kelle accelerator. 


\subsubsection{End-to-End Performance Improvement}
Figure \ref{fig:hw_eval1} compares the above baseline solutions in terms of energy efficiency and processing latency on multiple LLM and datasets.  On average, Kelle+eDRAM achieves a $3.94 \times$ and $4.46 \times$ improvement in latency and energy efficiency compared to the Original+SRAM, and performance gap gets larger as the decoding sequence gets longer. The superior performance of Kelle stems from the algorithmic innovations of AERP and 2DRP, combined with hardware advantages such as the efficient eDRAM memory controller, the systolic evictor design, and the proposed Kelle scheduler.


\subsubsection{Individual Contribution on the Performance Improvement}
In this section, we investigate the individual impact of Kelle optimization techniques. First, on average, compared to Original+SRAM, Original+eDRAM improves the speedup by $32\%$ but degrades the energy efficiency by $39\%$. The increased energy consumption is attributed to eDRAM refresh operations without algorithmic or hardware-level optimizations. 
The eDRAM enhances speedup due to its larger capacity and faster access speeds compared to SRAM. 
Second, the attention-based eviction policy accelerated by the systolic evictor reduces the latency by $2.39\times$ and improves the energy efficiency by $2.41\times$, when comparing the AEP+SRAM system with the Original+SRAM system. Next, thanks to the attention-based recomputation policy, AERP+SRAM system improves the speedup and energy efficiency by $1.19\times$ and $1.27\times$ to the AEP+SRAM system. 
Finally, for the system executing models with AERP, the eDRAM optimized by 2DRP and Kelle scheduler provides a $1.29 \times$ improvement in speedup and a $1.45 \times$ improvement in energy efficiency when comparing Kelle+eDRAM to AERP+SRAM system. Specifically, the 2DRP mechanism greatly reduces refresh energy, enabling the Kelle to take full advantage of eDRAM.

\subsubsection{Overhead Analysis}
\label{sec:overhead}
The pie charts in Figure \ref{fig:hw_eval1} display the energy breakdown for the Kelle+eDRAM system. 
The reduced energy share for the KV cache highlights how eDRAM coupled with Kelle algorithms alleviates memory access bottlenecks. Thanks to the efficiency of systolic arrays for matrix-matrix multiplication, the hardware overhead from KV recomputation is minimal, with RSA consuming only a small portion of on-chip energy.

To accelerate the token eviction process, we introduce a Systolic Evictor unit, which is a small computation unit coupled with the RSA. It takes an area of $0.06mm^2$ ($0.6\%$ of the on-chip area) and consumes power of 0.028W ($0.4\%$ of the on-chip power).  The systolic evictor avoids the stall of the LLM execution for KV cache eviction and redundant memory and computation access. It improves the system's energy efficiency and reduces the latency by $5\%$ and $7\%$.


\subsection{Comparison with Other Accelerators}
\label{sec:eval:accel_comp}

We compare Kelle+eDRAM with other cutting-edge LLM accelerators. 
LLM.npu~\cite{xu2024fastondevicellminference} enhances the on-device Neural Processing Unit (NPU) offloading to reduce pre-filling latency by re-constructing the prompt and model. DynaX~\cite{dynax} proposes dynamic fine-grained structured pruning to enhance the efficiency of sparse attention computation and achieves a $90\%$ attention sparsity. Dynax alleviates the computation bottleneck during the pre-filling stage. COMET~\cite{liu2024cometparticalw4a4kv4llms} quantizes the LLMs to 4-bit and designs high performance GPU kernel to support the mixed-precision computation. As advanced quantization techniques are not the main focus of this paper, we configure COMET to quantize LLM weights to 8-bit and both activations and KV vectors to 4-bit, ensuring a comparable KV cache storage budget to that of Kelle+eDRAM. 
Finally, we compare Kelle with the NVIDIA Jetson Orin edge GPU~\cite{nvidia-gpu} implementation of LLM using FP8, measured with pynvml~\cite{pynvml} and nvidia-smi~\cite{nvidia_smi}.

As shown in Figure \ref{fig:hw_eval_accelerators}, Kelle+eDRAM achieves better improvements in speedup and energy efficiency over other LLM accelerators. LLM.npu and Dynax optimize the computation-intensive pre-filling stage but do not address the KV cache bottleneck encountered during the LLM decoding stage. The performance gains of Kelle over COMET underscore the limitations of relying solely on KV cache compression without dedicated hardware accelerator support.

\subsection{Ablation Study}    
\subsubsection{Impact of KV Cache Budget}
\label{sec:kv_cache_ablation}
\begin{table}
    \caption{Energy efficiency over multiple KV cache budgets.}
    \centering
    \resizebox{0.85\columnwidth}{!}{%
       \begin{tabular}{c|c|c|c|c|c}
        \hline 
    N' in PG19 & 2048 & 3500 & 5250 & 7000 & 8750 \\ \hline 
    LLaMA3.2-3B & $8.07\times$ & $6.89\times$ & $5.77\times$ & $5.13\times$ & $4.55\times$  \\ 
    LLaMA2-13B & $5.06\times$ & $4.62\times$ & $4.02\times$ & $3.46\times$ & $3.11\times$  \\ 
    \hline 
    \end{tabular}
}
\label{tab:kv_budget}
\end{table}



Table~\ref{tab:kv_budget} illustrates the energy efficiency improvement of Kelle+eDRAM over different KV cache budgets $N'$. Without eviction, the largest possible number of tokens will be $N'=8750$ for PG19. Results indicate that even under this condition, Kelle achieves approximately \(3\times\) greater energy efficiency over Original+SRAM, highlighting the robustness of Kelle.

\subsubsection{Impact of the Recomputation}
\label{sec:eval_recomp}

\begin{figure}
    \centering
    \includegraphics[width=0.9\columnwidth]{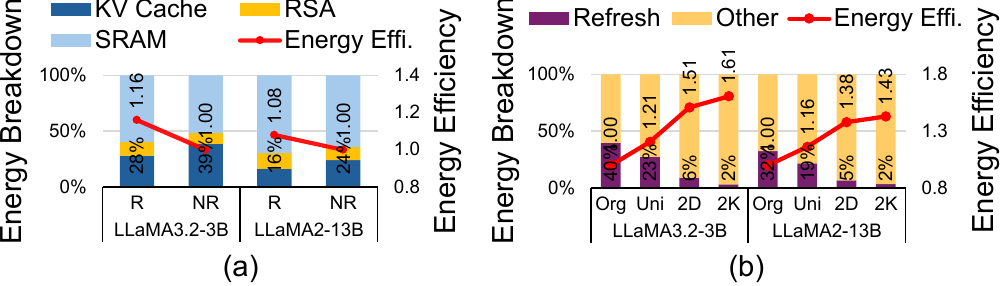}
    \caption{(a) Impact of KV cache recomputation in Kelle+eDRAM. 
    (b) Evaluation on 2DRP and Kelle scheduler.}
    \label{fig:ablation_2drp_recomp}
\end{figure}

\begin{figure}
    \centering
    \includegraphics[width=0.9\columnwidth]{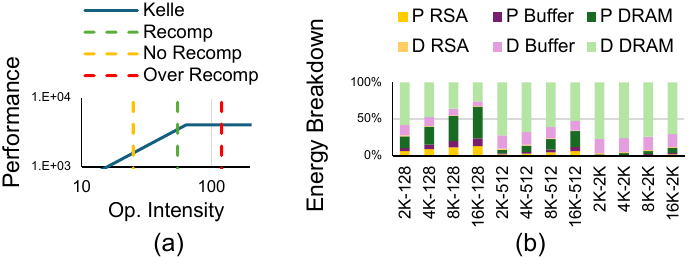}
    \caption{(a) KV Cache Recomputation impact. (b) Evaluation on long input sequences. P and D denote the prefilling and decoding stage, respectively.}
    \label{fig:prefill}
\end{figure}

We compare the energy consumption of Kelle+eDRAM with and without KV cache recomputation. As shown in Figure~\ref{fig:ablation_2drp_recomp} (a), the recomputation algorithm effectively reduces the energy consumption of KV cache, with minimal increase in RSA energy consumption. Moreover, we profile the popularity change of tokens during the pre-filling and decoding stage across LLM architectures and tasks. 
On average, over $86\%$ of the popular tokens in the pre-filling stage continue to be popular in the decoding stage, validating the execution strategy outlined in Section~\ref{sec:algo:recomp}.

Recomputation allows Kelle to store more tokens on-chip, reducing DRAM access. When processing the LLaMA2-7B model, accessing one KV vector from DRAM takes approximately $1.1\ \mu s$. In comparison, recomputing a KV vector using the RSA introduces an additional latency of $3.2\ \mu s$. Recomputation helps hide memory stalls by overlapping compute with memory access, reducing overall latency and improving energy efficiency by an average of $25\%$. For example, loading four KV vectors from DRAM requires $4.4\ \mu s$. With recomputation, three vectors are loaded, and one is recomputed in parallel during the load, reducing the total latency to $3.3\ \mu s$. In terms of energy, the RSA remains active regardless of the number of input vectors, so the incremental energy cost of recomputation is negligible. 

Figure~\ref{fig:prefill} (a) presents the roofline model of Kelle under three settings: No Recomp (without recomputation), Recomp (with a moderate recomputation workload), and Over Recomp (with excessive recomputation). Recomputation improves performance by increasing the effective memory bandwidth. However, as more KV vectors are recomputed, the RSA becomes a bottleneck. This behavior is reflected in the Over Recomp line, where Kelle transitions from a memory-bound regime to a compute-bound regime.

\subsubsection{Impact of 2DRP and Kelle Scheduler}
\label{sec:impact_refresh}

We evaluate Kelle+eDRAM running the Llama2-7B model executing the PG19 task under four strategies. \textbf{Org} strategy refreshes the eDRAM at its retention time with a $45 \mu s$ interval, ensuring almost no data corruption. \textbf{Uni} strategy uses a uniform refresh interval of $0.36ms$, the interval that enables the same LLM accuracy achieved by 2DRP.  2DRP, denoted as \textbf{2D} applies varying refresh intervals based on attention scores and bit positions. \textbf{2K} strategy combines both 2DRP and the Kelle scheduler. As depicted in Figure~\ref{fig:ablation_2drp_recomp} (b), the finer-grained refresh policy in 2DRP improves energy efficiency. With both 2DRP and Kelle scheduler, the Kelle achieves the optimal performance.

\subsubsection{Impact of eDRAM Retention Time} 


We assess the impact of eDRAM retention time on Kelle's performance, considering its influence on the bit failure rate. Retention time is affected by various factors, including design, technology nodes, and temperature~\cite{16nmEDRAM, edram_distribute, zhang2024camel}. We evaluate Kelle+eDRAM on the TriviaQA and PG19 tasks using 2DRP with different retention times. Specifically, we reduce the Kelle retention time (\(45 \mu s\)) to average refresh intervals of \(525 \mu s\), \(262 \mu s\), and \(131 \mu s\), respectively.
Table~\ref{tab:temp_ablation} shows the energy efficiency of these two settings compared to the Original+SRAM system. Thanks to AERP, the KV cache access overhead remains a small fraction of the total energy consumption. Consequently, the energy increase due to the retention time decrease is small, allowing Kelle+eDRAM to maintain performance gains. 

\subsubsection{Impact of Input Sequence Length}
\label{sec:eval_input_len}
We evaluate the energy consumption of Kelle+eDRAM under long input sequence lengths using the Llama2-7B model on the PG-19 dataset across different input output sequence lengths. We use the format input length - output length (e.g., "16K-128") to denote each experiment setting. 
As shown in Figure~\ref{fig:prefill} (b), when the input sequence is long and the decoding length is short, the prefilling stage dominates the overall energy consumption and the system becomes compute-bound. In this case, Kelle achieves a moderate energy efficiency improvement of $2.1\times$ over the Original+SRAM baseline. As both the input and output sequence lengths increase, the DRAM access energy for activations grows correspondingly. Under this more memory-intensive scenario, Kelle delivers an average energy efficiency improvement of $5.6\times$ over Original+SRAM and $1.8\times$ over AERP+SRAM, owing to its efficient KV cache management strategies.

\subsubsection{Impact of Batch Size}
\label{sec:batch_size}

\begin{table}[t]
  \centering
  \begin{minipage}[t]{0.42\columnwidth}
    \caption{Energy efficiency across  refresh intervals.}
    \centering
    \resizebox{\columnwidth}{!}{%
      \begin{tabular}{c|cc}
        \hline
        LLaMA3.2-3B & \multicolumn{2}{c}{Task} \\
        Interval ($\mu s$) & TriviaQA & PG19  \\
        \hline
        1050 & $3.91\times$ & $8.07\times$  \\
        525     & $3.65\times$ & $7.31\times$  \\
        131 & $3.06\times$ &  $6.05\times$ \\
        \hline
      \end{tabular}
    }
    \label{tab:temp_ablation}
  \end{minipage}
  \hfill
  \begin{minipage}[t]{0.52\columnwidth}
    \caption{Energy efficiency across different Batch Sizes.}
    \centering
    \resizebox{\columnwidth}{!}{%
      \begin{tabular}{c|cccc}
        \hline
        Task & \multicolumn{4}{c}{PG} \\
        \hline
        Batch & Original & AEP & AERP & Kelle \\
        size & +SRAM & +SRAM & +SRAM & +eDRAM \\
        \hline
        16 & $1\times$ & $3.16\times$ & $4.33\times$ & $6.67\times$ \\
         4 & $1\times$ & $1.71\times$ & $1.81\times$ & $2.23\times$ \\
         1 & $1\times$ & $1.24\times$ & $1.36\times$ & $1.71\times$ \\
        \hline
      \end{tabular}
    }
    \label{tab:eval_batch}
  \end{minipage}
\end{table}

We compare Kelle performance across different batch sizes using the Llama2-7B model on the PG-19 dataset, as shown in Table~\ref{tab:eval_batch}. While the energy efficiency improvement of Kelle over the Original+SRAM baseline is less significant at smaller batch sizes due to reduced utilization of the RSA and lower data transfer efficiency for model weights, Kelle still consistently outperforms all baselines. At a batch size of 1, Kelle achieves a speedup of $71\%$ over Original+SRAM, $37\%$ over AEP+SRAM, and $25\%$ over AERP+SRAM.

\subsubsection{Impact of eDRAM Bandwidth}
\label{sec:edram_bw}

We conduct experiments to evaluate Kelle under reduced eDRAM bandwidth (128GB/s), achieved by halving the number of banks and doubling the capacity per bank, while keeping the total eDRAM area and capacity constant. Using the Llama2-7B model on PG-19 and TriviaQA, Kelle+eDRAM achieves $1.47\times$ and $1.35\times$ energy gains over AERP+SRAM, and $6.31\times$ and $5.42\times$ over Original+SRAM. Though slightly lower than full-bandwidth Kelle, these results show that increasing eDRAM capacity, even with reduced bandwidth, effectively cuts costly DRAM accesses and improves bandwidth efficiency.

\subsection{Discussion}
\label{sec:discussion}
\subsubsection{Handle Long-Context Inference}
\label{sec:long_input}
    For long context inference, due to limited eDRAM capacity, excess KV data is offloaded to 16 GB DRAM. A simple analysis with LLaMA 2 7B shows that Kelle can support up to 19,000 input tokens without AERP, assuming 8-bit weights occupy 6.5 GB out of 16 GB DRAM and each token’s 16-bit KV pair across 32 layers. Introducing AERP enables immediate KV cache reduction after each layer's execution, freeing memory to accommodate the full input sequence in later layers. This allows Kelle to support input sequences of up to around 60K tokens. 
    Additionally, quantizing KV vectors to 4-bit enables support for up to 240K tokens. While an upper limit remains, it exceeds typical LLM input lengths up to tens of thousands of tokens~\cite{touvron2023llama,team2024qwen2,bai2025qwen2}.

Although longer input sequences increase overhead, the permutation invariant property of Equations~\ref{eqn:kv-computation1} and~\ref{eqn:kv-computation2} allows new KV vectors to be placed in the same positions as evicted ones, greatly simplifying the paging process. Additionally, 
the vectors can be prefetched sequentially without requiring complex lookup mechanisms. As a result, the prefetching overhead increases linearly with the input length, avoiding disproportionate growth.

\subsubsection{Integrate Kelle with GPU}
\label{sec:gpu_integration}
While Kelle is implemented with a systolic array, AERP can be adapted to GPUs; however, identifying the token with the lowest attention score may be inefficient due to the lack of a systolic evictor. 2DRP is specific to eDRAM to reduce the refresh energy consumption. The eDRAM can be coupled with the GPU’s existing memory system to store KV vectors. 
Finally, the Kelle scheduler can be readily implemented using CUDA.


\section{Conclusion}
\label{sec:conclusion}
The KV caching technique is crucial for enhancing the efficiency of LLMs. However, storing the extensive KV vectors results in a substantial memory footprint and increased data access costs. In this work, we introduce Kelle system that utilizes eDRAM as the primary storage medium for KV vectors. The superior performance of Kelle highlights the significant potential of eDRAM in implementing the KV caching mechanism, paving the way for future research.


\bibliographystyle{ACM-Reference-Format}
\bibliography{sample-base}

\end{document}